# A Quantum Matter-Photonic Base Framework for Describing Attosecond Physicochemical Processes

O. Tapia
Chemistry-Ångström
Uppsala University, Box 259,
75120 Uppsala, Sweden
email: orlando.tapia@fki.uu.se

**Abstract**
Prerequisites mandatory to fulfill headline's program imply recasting several basic concepts: 1) Different perception of quantum states; sequence of maps from abstract Hilbert space down to laboratory levels via inertial frames from Special Relativity. 2) Definition of multi-partite basis allowing for chemical tenets via linear superposition principle; multi-partite space support coherent states and decoherence. 3) Fock space (photon number) and concomitant photonic bases tying together matter-to-photon quantum states; 4) Photonic base state units. 5) Feshbach-like resonance states: continuum-to-discrete multi-partite base states coupling. 6) Electro-nuclear base states allowing for natural inclusion of microwaves and radio frequency (rf) radiation, the former enters not only as heating sources but also as foundations of catalytic activity; rf-radiation as propagation regulators of electro-nuclear quantum states.

## Introduction

This work analyzes a framework to help describing processes formulated in quantum physical terms beyond semi-classic schemes thereby debasing features of representational modes such as potential energy surfaces, particle models, etc.[1-9] The result is a quantum scheme where the mathematics of Hilbert space remains untouched but foundational quantum tenets differ from standard ones.

Theoretical studies in attosecond physics and chemistry require a full-fledged quantum physical framework including light-field-controlled pulses; this is one hypothesis lying behind the present work.
At laboratory level electromagnetic (EM) radiation and matter interaction discloses its quantum nature. Planck (1900) linked EM radiation frequency to energy exchanged with an action constant, the famous Planck constant.
The framework would help construct theoretically consistent views once introduction of photonic bases complete the standard quantum formalism. This leads to non-separable views as photon and matter fields are tied to each other. The concept of quantum states occupies center stage.

Basis sets construction can proceed by focusing on electronic processes relevant to chemical physics as well as material sciences, biophysical and astrophysical processes. A quantum framework obtains where classic particle/wave duality plays no fundamental role; only a pedagogic one to eventually help beginners initiating into an otherwise unusual and abstract world of quantum (q)-states.
Electron-q-states control, at attosecond time scale, concerns both experiment and theory. The scheme is potentially useful to handle e.g. photonic energy confinement and transformation via chemical paths.

The photonic scheme permits including femtosecond (and lower) time scale phenomena [1] yielding a unified view.

One encounters separation problems that are related to four sorts of entanglement involving: 1) electronic states, ES; 2) nuclear spin states, NSS; 3) photon-matter states, PMS; 4) photon states, PS.

If we keep in mind that a classical world makes part of (is embodied by) a broader quantum physical one, many classical techniques that we use today, properly rigged,[6] would play subsidiary roles in opening new vistas over old and new fields of great interest not only to Physics and Chemistry but also Biology and Technology. No reductionism implied.

Classical physics elements are identified by mappings between abstract and laboratory space elements, basically introducing inertial frames (I-frames) from special relativity theory (SRT) and transformation (invariance) groups; these elements are hence merged in quantum frameworks with algebraic (graded) models and suitable algorithms helping calculate base states sustained by electro-nuclear (EN) elements;[6] *information* so gleaned re-injects as quantum labels; abstract-like basis set obtain, and linear super-position principle properly reinstated helps presenting processes, not representing them.[3-6]

Entanglement and coherence [10] remain quantum physical phenomena holding a key to apprehend the nature not only of chemical bond; coherent energy transfer in biological systems is another case of interest concerning quantum states. With focus shifted away from the material sustaining them, coherent states turn out to play central roles in rationalizing not only attosecond chemistry, of course.

Time and space are central concepts. Quantum entanglement raises important queries on locality that imposes an attentive perception of configuration space framework. On the one hand, abstract quantum states should handle all response-possibilities available to physical systems. On the other hand, probabilities emerge from laboratory demolition measurements, namely, *events* at the fence of two worlds.[5,6] These events in the present framework would convey information richer than just counting sequences. [5]

Thus, a functional view replaces the standard representational mode of quantum mechanics; as a consequence, weirdness accusations are left behind.

The paper in six sections presents indicative paths we think required to attain a fuller quantum physical framework compatible with SRT.[2,3,6-9,11-14] First, Time and Space prolegomena set up configuration space. Second, multipartite basis set concept and entanglement of q-states sustained by material elements is examined; here one aims at finding bridges to quantum probing. Third, presents standard quantum formalisms (scattering scheme and probing of wave functions) to underline mappings from abstract to laboratory levels.

Fourth stage introduces a photonic scheme; base states are defined, together with a minimal photon-matter base state unit. These latter appear associated to sets of amplitudes; they are not objects, though, sustained by well-defined elementary materials.

Coherent states and decoherence are defined and used to discuss q-processes. Q-states presented as linear superpositions over photonic base states [1] are expected to help apprehend chemical meanings beginning from a full fledged quantum perspective;[7-9] moreover, relations to mechanistic issues involving processes having physical, chemical and/or biological import appear in a new light.

Fifth stage introduces generic two-state frameworks basis adapted to photonic base states. A connection with computing schemes becomes possible though not explored here. The role of high intensity electromagnetic fields is brought to attention.

Sixth, attosecond processes in an electronuclear (EN) photonic basis scheme make up for the following section.

A general discussion closes the paper; some key QM issues discussed here underline the difference with standard views.

**Time and space: prolegomena**

Time and space are foundational concepts. This means that one cannot define them in absolute terms. Yet perception is phenomenological; this fact helps measurement on quantitative scales, e.g. femto-, atto- seconds, hours, days, years, etc. Thus one can get a measure of Time via *intervals*, and a measure of Space via other kinds of *intervals*; relative quantities can be probed and therefore are measurable.

Definitions of foundational concepts are better avoided then. The preferable approach defines measures related to them. For us, it is a task accomplished by Special Relativity Theory (SRT) that provides mathematics allowing us to think and work within well-defined contexts. In particular the idea of absolute time and space is excluded.

**Space-time framework**

First relate (map) q-states belonging to the abstract level with those in laboratory surroundings recurring to scalar product mappings.

SRT introduces 4-dimensional (4D) space-time frames, including 3D-inertial (I)-frames and a special time t-axis. As long as one is constructing adequate basis sets such frames are not submitted to (classic) accelerations; this is an important caveat.

It comes as no surprise that the concept of q-state is foundational too; so we use mathematical elements related to abstract quantum states. These are elements of linear vector spaces over the field of complex numbers.

I-frames, besides letting in a special time-axis (t-), permit introducing a configuration space, e.g.: $\mathbf{x} = (\mathbf{x}_1,\ldots,\mathbf{x}_n)$ with the number 3n equaling the *total number* of classic degrees of freedom; and $\mathbf{k}=(\mathbf{k}_1,\ldots,\mathbf{k}_n)$ corresponds to a reciprocal configuration space. These numbers are used as label to Dirac's kets; e.g. configuration space vectors: $\{|\mathbf{x}>=\{|\mathbf{x}_1,\ldots,\mathbf{x}_n>\}$; and its reciprocal space vectors: $\{|\mathbf{k}>=\{|\mathbf{k}_1,\ldots,\mathbf{k}_n>\}$. No particle coordinates intended, just abstract mathematical spaces acting as supports to functions as shown later on. Quantum degrees of freedom are introduced later on with the help of quantum numbers.

The amplitudes (scalar products) $<\mathbf{k}|\mathbf{x}>$ and $<\mathbf{x}|\mathbf{k}>$ play the role of base functions: $\exp(i\mathbf{x}\cdot\mathbf{k})$ and $\exp(-i\mathbf{x}\cdot\mathbf{k})$, respectively; Fourier transforms connect q-states in a $\mathbf{k}$-basis $g_\mathbf{x}(\mathbf{k})$ to one in an $\mathbf{x}$-basis $g_\mathbf{k}(\mathbf{x})$ and vice-versa;

these linear superpositions are usually named wave packets; it is a particular weight function that defines the corresponding quantum state.

Note that the I-frame fixes an origin (relative to another I-frame) to the configuration and reciprocal spaces so that the quantum states would share a uniform state of motion (if any) conveyed by I-frames; this is a map chosen between quantum and classical domains; a boundary is introduced here.

Now, the sets {|**x**>} and {|**k**>} of basis kets are given the structure of rigged Hilbert spaces via introduction of generalized functions such as Dirac's delta ($\delta$) and its derivatives:[6,17] e.g., <**x**|**x**'>→$\delta$(**x**-**x**') and <**k**|**k**'>→$\delta$(**k**-**k**').

A similar situation holds for the conjugate magnitudes concerning time and frequency that, when *used as labels*, leads to base sets |t> and |$\omega$>; thus, exp(i$\omega$t) and exp(-i$\omega$t) correspond to <$\omega$|t>= exp(i$\omega$t) and <t|$\omega$>= exp(-i$\omega$t); these scalar products play the role of basis functions for the time-frequency connections via Fourier transforms. These base transformations let relate quantum states in $\omega$-space to quantum states in t-space, and vice-versa.

Suppression of a particle representation mode dismisses criticisms concerning planar "waves" as base sets.

One has to get at laboratory level to endow the ($\omega$, t) numbers with physical relevance; e.g., electromagnetic energy ($\hbar\omega$) provides support for quantized basis set when expressed as quanta numbers: $n_\omega$=0,1,2…Bridging abstract to laboratory elements is done with Bohr's postulate: transition energy between two (matter-sustained) base states relates to electromagnetic energy.

A measure of energy defines what is physically meaningful as is the case for space and time. Thus, base states {|t>} and {|$\omega$>} do not represent "time" or "frequency" at this stage where construction of basic elements is the matter, t-symbol stands for parametric time as the one featuring Schrödinger equation. One has to introduce appropriate measures to proceed with laboratory time.

In practice, distinction of abstract mathematical schemes and projected ones is gone if one renounces to the idea that particles occupy energy levels; all weirdness associated to quantum mechanics when described in classical terms dissipates.

*No need for an interpretation of* QM with classical physics pictures; q-states sustained by particles, yet not giving them a representation.

Time is parametric in abstract and physical q-states, though physical probe (demolition-measurement) links q-states to laboratory time and energy-momentum conservation principles (see below).

It is common practice to take $\hbar$=1 to transform the symbol $\omega$ into an energy *pulse*.[15] In conventional terms, exp(-i$E_\omega$t/$\hbar$) and its complex conjugate permit *transforming basis* from energy to time dimensions. Note, *neither energy nor frequency are quantized in themselves*; what is quantized corresponds to *energy exchanged* between quantum systems. When this latter is an EM field, the energy quantum is named photon: unit of exchanged energy. A similar situation holds for angular and linear momentum probes.

Under less fine-tuned experimental conditions wave packets ($g_\omega(t)$ or $g_t(\omega)$) are produced with finite frequency width and height. If such packet associates a central frequency, the corresponding energy (eigenvalue) is said to show a finite lifetime. Femtosecond and attosecond pulses are conspicuous cases.

Thus, Bohr's *postulate* puts up a mapping between Hilbert space *energy gaps* with a frequency $\omega$ attached to q-EM field and qualifications just presented. Selection rules follow e.g. from angular momentum conservation laws. [6,17]

Most important: Energy levels do not represent energy of the materiality; they are very useful mathematical tools (usually reflecting boundary conditions). The actual numeric value is defined to within a constant, e.g., for an I-frame in uniform motion, kinetic energy with respect to a second I-frame locating a receptor device to be used to detect the emission; the relative kinetic energy would shift the whole spectra (z-shift in astronomical spectroscopy [18]). The information carrier is also a quantum state: e.g. a photon state (wave packet) or a quantum state sustained by matter aggregates.

The symbol $\langle x|\Psi,t\rangle = \langle x_1,\ldots,x_n|\Psi,t\rangle$ stands for abstract quantum state projection on (abstract) configuration space; it corresponds to a scalar product, a complex number or complex function $\Psi(\mathbf{x},t)$ over the field of real numbers $\mathbf{x}$ and $t$; these number fields are the functions' support. These functions *are not* the mathematical entities assumed at dawn of quantum theory by Schrödinger; what is not retained here is the *representational* character via particle position space. In a more careful presentation one should identify the time-axis origin ($t_o$) by specifying $\Psi(\mathbf{x}, t_o)$. Only relative time evolution makes sense.

A quantum state can be projected on a rigged Hilbert space. Finally, group invariances enter via I-frames group transformations [6,17].

**Time & Space in Laboratory Setups**

Light pulses help measuring short time intervals as well as perceiving dynamic features. The femtosecond barrier was broken via Fourier synthesizes to generate pulses of few attosecs (1as= $10^{-18}$ sec).

The nature of this discovery and recent developments will shake views on processes both physical and chemical.

We will refer and use some cases of attosecond pulses within a quantum physical chemistry framework.

Observe that a wave function such as $\Psi(\mathbf{x}, t)$ blurs an important issue: the state of motion associated to the corresponding I-frame. The need for a second I-frame is consubstantial to SRT, and this latter is used to pinpoint the phase change associated to this wavefunction: $\exp(iG(\mathbf{R}(\tau)))$ where $\mathbf{R}(\tau)$ is a real space point measured from a second I-frame at time ($\tau$) in that frame. Laboratory time would enter into the scheme via such a phase and can be indicated with Greek letter $\tau$. Both times are independent unless special interactions between internal (t-space) and external ($\tau$- space) are set up.

## Multi-partite General Basis Sets

The analysis of attosecond processes is performed from high-energy domain down to low energy regimes. In this context, concepts of molecules or super molecular fragments thereof are too rigid to handle (label) quantum states in abstract terms; the *partite* concept as introduced below is retained by its flexibility.

Spectral responses distinguish different types of q-states. Even if the same number of elementary constituents sustains all q-states they are differentiated by quantum number groupings and particular responses. To analyze them properly there is need for basis sets types identifying multi-partite state elements; such partite-states when taken separately may refer to fully autonomous spectral sources. In the present context they are not to be seen as objects in a classical sense unless one can associate an independent I-frame to quantum states that are sustained by the materiality whose number of elementary terms remain fix.

The form of a generic basis sets is given below as:

i) $\langle x_1,\ldots,x_m | \phi_{k1\ldots km} \rangle$ (1-1)

ii) $\langle x_1,\ldots,x_{m-1} | \phi_{k1\ldots km-1} \rangle \otimes$
$\langle x_m | \phi_{km} \rangle$ (1-2)

iii) $\langle x_1,\ldots,x_{m-2} | \phi_{k1\ldots km-2} \rangle \otimes$
$\langle x_{m-1} | \phi_{km-1} \rangle \otimes \langle x_m | \phi_{km} \rangle$ (1-3)

…

m) $\langle x_1 | \phi_{k1} \rangle \otimes \langle x_2 | \phi_{k2} \rangle \otimes \ldots \otimes$
$\langle x_{m-1} | \phi_{km-1} \rangle \otimes \langle x_m | \phi_{km} \rangle$ (1-m)

The bases are referred to as one-partite (1-1), bi-partite (1-2), tri-partite (1-3) till m-partite (1-m); quantum numbers are variable identifying the bases. A partite family of base sets is reckoned from the number of direct product symbols plus one to the extent these latter separate different quantum number groups. The set of quantum numbers $\{k_1\ldots k_m\}$ identifies the base functions.

Presentation of general quantum states require infinite dimension base vectors that would read in an economic format:

$(|1\text{-}1\rangle \ldots |1\text{-}2\rangle \ldots |1\text{-}3\rangle \ldots |1\text{-}m\rangle \ldots)$ (2)

And in an extended mode with symbol **x** indicating product operation:

$(\phi_{k1\ldots km} \ldots \phi_{k1\ldots km-1} \mathbf{x} \phi_{km} \ldots \phi_{k1\ldots km-2} \mathbf{x} \phi_{km-1} \mathbf{x} \phi_{km} \ldots \{\phi_{k1} \mathbf{x} \ldots \mathbf{x_{m-1}} \phi_{km}\})$ (3)

This base set is sufficiently general. Thence, an arbitrary quantum state is given as the transpose vector collecting ordered (particular) amplitudes:

$(C_{k1\ldots km} \ldots C_{k1\ldots km-1} \mathbf{x} C_{km} \ldots C_{k1\ldots km-2} \mathbf{x} C_{km-1} \mathbf{x} C_{km} \ldots \{C_{k1} \mathbf{x} \ldots \mathbf{x_{m-1}} C_{km}\})^T \rightarrow$ (4)
$(C_{1\text{-partite}} \ldots C_{2\text{-partite}} \ldots C_{3\text{-partite}} \ldots C_{m\text{-partite}})^T$

Forms (2) or (3) imply all base states enter and remain fixed. While anything happening to a given quantum state, e.g. a particular change, the amplitudes are to mirror the change in eq.(4).

Note that a quantum state *is not* an object or a representation of particles.

In a way, linear superpositions over this type of basis set may express all the chemistry sustained by a fixed materiality; a careful specification of degrees of freedom (quantum and classic) is required however before considering specific cases; e.g. spin angular momentum permits getting a more faithful presentation.

Symbolically, the most general q-state |Φ> reads as the scalar product:

$$|\Phi\rangle \rightarrow (\phi_{1\text{-}P} \ldots \phi_{2\text{-}P} \ldots \phi_{3\text{-}P} \ldots \phi_{m\text{-}P}) \bullet (C_{1\text{-}P}(\Phi) \ldots C_{2\text{-}P}(\Phi) \ldots C_{3\text{-}P}(\Phi) \ldots C_{m\text{-}P}(\Phi))^T \quad (5)$$

In the business of *basis set construction* there is no interactions among partitioned sets; no particle representation attempted.

Firstly, one proceeds to formally construct all possibilities open to the system; then, once the system under study is secured, appropriate operators incorporate interactions to determine amplitudes for particular cases. They stand either as internal operators or operators for external sources (mixed).

Among partite basis, one can differentiate two families:

1) All partition bases are referred to one and same I-frame, only quantum numbers are involved in identifications (labeling) including those associated to the I-frame (e.g. rotations and translation invariances); [2-6]

2) Different partitions are assigned to different I-frames albeit for some cases conserve a master I-frame (box). These I-frames permit defining real space distance maps, relative orientations and relative time measures. Absolute time and space fades away.

For case 2) one is preparing for a typical laboratory situation with varied alternatives defined by experimenters, namely, where and how to introduce detectors, double slits, mirrors, phase modulators, beam splitters, etc.

When all partitioned base states share the same I-frame then, besides internal quantum numbers, they can also be globally label with quantum numbers coming from a common box (e.g., equal box-lengths, and different material parameters); this labeling adds on top of the (internal) partite quantum numbers; it is supernumerary. [3-6]

In summary: box-quantum-numbers provide an "external" characterization to the I-frame material system states, as a whole, while "internal" quantum numbers are sustained on the configuration space associated to each partitioning [2-4] including spin.

These two classes and subclasses of quantum numbers are needed to describe laboratory processes via the linear superposition principle; coherence and decoherence are among them. The invariant element is the total number of elementary material constituents found in the trapping device (volume V container and possibly pressure and thermal bath).

**Entanglement and Probing**

Entanglement expresses a quantum foundational concept too. It concerns quantum states as such and not particles sustaining them (see below).

Construct the states $|\pm\rangle_{12}$ given as coherent linear superpositions for a bipartite system:

$$|\pm\rangle_{12} = [(1/\sqrt{2})\langle\mathbf{x}_1|\phi_{k'1}\rangle \otimes \langle\mathbf{x}_2|\phi_{k'2}\rangle \\ \pm (1/\sqrt{2})\langle\mathbf{x}_1|\phi_{k'2}\rangle \otimes \langle\mathbf{x}_2|\phi_{k'1}\rangle] \quad (6)$$

This bipartite element stands for two entangled states, $|+\rangle_{12}$ and $|-\rangle_{12}$.

As it is written $|\pm\rangle_{12}$ corresponds to two non-separable (and orthogonal) modes. Now include the remaining degrees of freedom and take this fixed linear superposition as element of the base set to get:

$$|\pm\rangle = |\pm\rangle_{12} \otimes \langle\mathbf{x}_3,\ldots,\mathbf{x}_m|\phi_{k'3\ldots k'm}\rangle \quad (7)$$

Note |±⟩ is a separable state with respect to |±⟩$_{12}$ and ⟨$x_3$,…,$x_m$|φ$_{k'3…k'm}$⟩; namely a direct product state. Yet it is a partially entangled state via |±⟩$_{12}$.

Let a probing device act by inducing a transition relating states |+⟩ to state |-⟩ at a given space location be it a scattering source, or detector/register, etc.; thus the resultant quantum state might yield either $(1/\sqrt{2})(|+⟩+|-⟩)$ or $(1/\sqrt{2})(|+⟩-|-⟩)$. This case expressed as change of base states reads:

Transition $T^+$: $(1/\sqrt{2})(|+⟩+|-⟩)$ → ⟨$x_1$|φ$_{k1}$⟩⊗⟨$x_2$|φ$_{k2}$⟩⊗⟨$x_3$,…,$x_m$|φ$_{k3…km}$⟩
(8a)

Transition $T^-$: $(1/\sqrt{2})(|+⟩-|-⟩)$ → ⟨$x_1$|φ$_{k2}$⟩⊗⟨$x_2$|φ$_{k1}$⟩⊗⟨$x_3$,…,$x_m$|φ$_{k3…km}$⟩
(8b)

These are possibilities accessible to the system interacting with real probing devices (they can be presented as amplitude changes in a fix base set).

The physical effect of probing, if successful, will *move amplitudes* from the entangled base component |±⟩$_{12}$ to a non-entangled one i.e. (8a) or (8b). The entangled state is hence "destroyed" while materiality remains untouched.

*Probing (measuring) changes number of partite elements by one; in this case increases by one. This is commensurate with e.g. a dissociation process when referred to a 1-partite state.*

Examine the situation in more detail:
Let partite base state eq.(7) serve as reference I-frame in uniform state of motion (conserved quantity); associate I-frames to each partite element in the direct product ⟨$x_1$|φ$_{k'1}$⟩⊗⟨$x_2$|φ$_{k'2}$⟩ that in **k**-space the I-frames displace in opposite directions: $\widehat{k}_1 = -\widehat{k}_2$, no matters internal quantum numbers: ⟨$k_1$|φ$_{k1}$⟩⊗⟨$k_2$|φ$_{k2}$⟩ & ⟨$k_1$|φ$_{k2}$⟩⊗⟨$k_2$|φ$_{k1}$⟩ are equivalent. Moreover, with respect to ⟨$x_3$,…,$x_m$|φ$_{k'3…k'm}$⟩, i.e. (m-2)-partite state that serves as anchor, their locations are opposite (antipodes); this results from conservation of angular momentum; I-frames axis orientation is then arbitrary if the outgoing state is an S state (spherically symmetric). All these statements refer to *accessible possibilities*. Not to particles.

Now, at any two correlated points in space locating detectors for instance, a response rooted at ⟨$x_1$|φ$_{k'1}$⟩ and the other at ⟨$x_2$|φ$_{k'2}$⟩, would click simultaneously as they belong to an entangled state; similarly for case ⟨$x_1$|φ$_{k'2}$⟩ and ⟨$x_2$|φ$_{k'1}$⟩. In a sense, linear and angular momentum conservation elicited by base states tie materiality location (*so to speak*) as materiality sustains the quantum state; you cannot get one without the other.

What happens if only one detector is used?

We are now near Einstein-Podolski-Rosen (EPR) paradox,[19] although with a fundamental difference (see also [5]). For, assembling only one sensor, that is an object you command from outside, a local probing of the entangled state with effective detections, namely, detecting there a response rooted say at ⟨$x_1$|φ$_{k'1}$⟩ implies that at the antipode there must be a "virtual" response to be rooted at state ⟨$x_2$|φ$_{k'2}$⟩ *whether you measure it or not*. If you, as EPR did, understand all these in terms of (independent) particles there seems to be a "spooky" action at a distance. And all adjectives on the weirdness of quantum mechanics "naturally" appear in describing events.

But a problem here is that in this kind of talking one mixes possibilities with actualities: this is a misunderstanding. One has to analyze the former first and thereafter set up the probing systems (in real space) and select possibilities *compatible under the experimental constraints*. And reminding that it is to quantum states that probing addresses and not to objects (electrons, atoms).

Let come back to the generic quantum state to be probed. All three elements first share the I-frame:

$$[1/\sqrt{2} \langle \mathbf{x}_1 | \phi_{k'1} \rangle \otimes \langle \mathbf{x}_2 | \phi_{k'2} \rangle \pm 1/\sqrt{2} \langle \mathbf{x}_1 | \phi_{k'2} \rangle \otimes \langle \mathbf{x}_2 | \phi_{k'1} \rangle] \otimes \langle \mathbf{x}_3,\ldots,\mathbf{x}_m | \phi_{k'3\ldots k'm} \rangle \quad (6)$$

Consider some points worth reminding first:
(i) Once entanglement sets up, quantum state eq.(7) is the same at any configuration space point as well as (parametric) time; ii) the quantum state under laboratory conditions is sustained by the elementary materiality; iii) Even if no materiality is present, the state remains to the extent this is an *abstract* state (a possible state) also, *no actual response for so long materiality is absent*. This latter point may seem strange. But, quantum physics is about possibilities and one of them may be absence in the volume of the sustaining material (vacuum). This is no more complicated than saying: whatever is done, no real physical response from the system rooted at the given quantum state is expected if no materiality shows up there to actually sustain it; this is what *presence* refers to. *This is a property of laboratory physical states under probing*.

Thus, materiality's presence is a key to physical responsiveness. [5,6]

Abstract quantum theory does not describe materiality whereabouts in real space, thence no trajectories. But it must be present at laboratory level to ensure conservation laws once for instance a $T^+$ probing successfully takes place. The one way (and possibly the only one in absence of noise) to get a response at detector is to realize the interaction with *direct intervention of the materiality sustaining the quantum state*.

At this point, and only at this one, I-frames over the partite states in a manner of speaking may become apparent; *this is a result of (real) interaction* (see Fence concept in Refs.[5,6]); it cannot happen in abstract Hilbert space.

So long entangled states evolve in *abstract* space all possibilities are accessible and can be calculated. But, for the present case, entanglement means that once materiality is detected, the result (detection) implies that both I-frames are unraveled by $T^\pm$ successful interactions. The $T^\pm$ transition connects spaces with different number of I-frames and might activate clocking; an energy exchange between the detector and the physical quantum state characterizes the probing. Parametric time leaves space for laboratory time.

Interestingly, this is what characterizes decoherence. Here lies one of quantum physics conundrum if a classical physics viewpoint is used to describe it.

The entangled quantum state has a special quality then: either it expresses as possibility or does it as presence *concomitantly* with probing realizing a particular event; making it possible.

If only one detector is present and if it were activated at a given laboratory time, the nature of the interaction, implying presence of the entangled state, would be triggered if and only if the second materiality component is present too, simultaneously yet not measured; otherwise, detection could not be actualized. This *information* comes from the entangled state.

There is no signal requirement between the terms of the entanglement before probing, no spooky action at a distance; actually there is no distance. All materiality implied by the entangled state must be "at the right place" at the "right time", the procedure amounts to prepare the quantum state for the non-probed partite state. This is the meaning under our concept: *quantum-state-sustained-by-a-given materiality*.

If one looks at the semi-classic case discussed by EPR it follows a simple result: for them, both "particles" are present but knowledge of their state is missing. Detecting the state for one of them the second is enforced at a distance in real space that may be beyond what is allowed by SRT. From the viewpoint used by us there is no such enforcement; *the state was already there*.

The quantum formalism imposes simultaneous action; yet there is no real space involved and no real time spent. Such is the nature of quantum entanglement. *It provides the experimenter with a resource*. A resource found at the grounds that sustain the idea of quantum computers as well as teleportation of quantum states (not particles, of course).

The problem with classical analyses is that it separates the entanglement terms and treats them independently of each other. Moreover, quantum physics does not address particle's motion; *I-frame motion* is the classical link. From the preceding descriptions one can conclude that: tenets of classical physics are not even wrong in quantum-entangled circumstances; they are irrelevant at best. Thus:

*Quantum entanglement cannot be simulated in classical physics terms*.

This conclusion closes a first view of quantum entanglement from the perspective developed here and in other publications.[1,5,6] It agrees with Bell's 1964 theorem, stating that the predictions of quantum theory cannot be accounted for by any local theory; this represents one of the most profound developments in the foundations of physics.[19] The agreement of the present approach with this important theorem is rewarding. Note that our approach concerns *not only* event counting but also and foremost amplitude's sensing. In other words, measuring wave functions would be a requirement.

**Multi-partite dissociation reactions**

Starting from one-partite state, probing partite response implies non-zero amplitudes outside the 1-partite domain. A pre-dissociation state reads for example as:

|pre-Dissociation> →
$(C_{1-P}(\Phi)...C_{2-P}(\Phi)...0_{3-P}(\Phi)...0_{m-P}(\Phi)...)^T$
(9)

The eq. translates into generic linear superpositions over two partite domains. This generic formulation cover all possibilities associated not only to 1- and 2-partite elements.

Quantum descriptions of chemical processes boil down to calculate amplitude changes connecting different partite domains.[4]

In particular, state vectors cover dissociated states when several I-frames can be in action; example of this is stated by eq.(10') below:

|Dissociation> →
$(0_{1-P}(\Phi)...C_{2-P}(\Phi)...0_{3-P}(\Phi)...0_{m-P}(\Phi)...)^T$
(10')

The I-frame for the one-partite states is used as reference and put there zero amplitude, namely $0_{1-P}(\Phi)$.

The state |pre-Dissociation> stands for a coherent state, only one I-frame for a number of partite basis states. Decoherence breaks this state of affairs so that I-frames for fragments can be seen moving with respect to the origin at a classical (relative) speed **v**. This is one result of introducing SRT inertial frames. The distance concept in space and time now enter more naturally. Yet they do not apply to abstract elements of a quantum framework; in a basis set construction there is no interaction between internal to external degrees of freedom. Appropriate operators are required for process descriptions.

Let introduce short time windows, that is time distance $\Delta t = t-t_o$ when an I-frame displaces a distance **ΔR** from the reference one; the relative velocity **v**=**ΔR**/Δt can serve as a label of base states sustained by the corresponding "mobile" I-frame; so called box states. The mass sustaining the internal quantum states is invariant leaving an opportunity to include classical dynamics "representation" elements (e.g. Lagrangians).

Now, put at the fixed frame a source and a detector at moving frame endowed with relative velocity **v**; from the frame in motion the initial frame is receding with -**v**. We have basic elements entering a measurement device in laboratory space. Because one is not in a representational mode one has to be careful when mapping quantum states. The I-frames become apparent only when an interaction changing the q-state with respect to the surroundings expresses by energy, linear, angular momentum changes. Such changes may produce an imprint in the detection device.

At this stage operators describing specific interactions enter. How do wave functions change due to such effects that might characterize probing devices?
Two generic situations are discussed below.

**Scattering & Wavefunctions**

Within the standard scattering formalism there are simple schemes to formulate detection situations [6,17]. Below the multipartite scheme is cloaked in scattering theory formalism. Take the quantum state given by eq.(5) (|Φ,t>) and assume that a base state written as |ϕ$_{k'}$> corresponds to a particular state belonging for example to the bi-partite sector.
Let C$_{k'}$(t) be the amplitude within the quantum state |Φ,t>. By construction at initial time C$_{k'}$(t$_o$)=0 implying that the system was prepared as 1-partite domain. That is C$_{1P}$(Φ,t$_o$)≠0, C$_{2P}$(Φ,t$_o$) = 0,..., C$_{mP}$(Φ,t$_o$)=0. At this time switch the interaction; clocks are running at laboratory premises.

After time lapse ($t_f-t_o$) a probe directed to get a response from base state $|\phi_{k'}\rangle$ is expressed as [6,17]:

$$T_{k'}(t) \equiv \langle \phi_{k'} | \Phi(t) \rangle \quad (11)$$

This amplitude is nothing else than $C_{k'}(t)$ understood the experiment starts at $t_o$. From scattering framework including $\hat{V}(t)$ scattering operator the amplitude is also written as:

$$T_{k'}(t) = (1/i\hbar) \int^t dt' \langle \phi_{k'} | \hat{V}(t')\Phi(t') \rangle \quad (12)$$

For k' base states belonging to a bi-partite domain the amplitude $C_{k'}(t)$ is giving us all information there is. We can use eq.(11) to actually point at data to be collected determined by the operator $\hat{V}(t)$ selected to stand for a measuring device.

But we have more information via relative I-frames. To each one of those assign direction **k**-labels. Consider a one-I-frame ("particle") case:

$\phi_{k'} \to |\mathbf{k'}\rangle$ and

$T_{k'}(t) \to \langle \mathbf{k'}|\Phi(t)\rangle \to$ Fourier-transform:
$\Phi_{FT}(\mathbf{k'},t) = g_x(\mathbf{k'},t)$

This is the Fourier transform of the exact spatial scattering state.

There is nothing paradoxical in projecting on a complete plane wave basis set. We are not projecting particles; in fact, there is no attempt at representing the material system (at all). Quantum physics addresses the possibilities accessible to the system than could experimentally be probed; actualities have to be handled on a case-by-case base. In particular, if a measuring device is located in real space at point **R**, then there is a weighting function to be included:

$\delta(\mathbf{r}-\mathbf{R}) = \langle \mathbf{r}|\mathbf{R}\rangle$.

If the detector position is given as **R**=**v**t, where t is the time lapse the I-frame takes to reach the detector with velocity **v**= $\hbar\mathbf{k}/\mu$ where effective mass $\mu$ is associated to the I-frame system one gets: [20]

$$G_k(x',t-t_o) = \left[\left(\exp(i\hbar k^2(t-t_o)/2\mu)\right)/(2\pi)^{3/2}\right] \times$$

$$\int dk' \Phi_{FT}(k')\exp\left(i(\hbar(t-t_o)/2\mu)(k'-k)^2\right) \quad (13)$$

Inclusion of inertial frames associated to the partite set opens the way to get an equivalent of the *imaging theorem*. Mutatis mutandis, the treatment of Briggs & Feagin [20] can be adapted to get a mapping to detectors characters in laboratory space. Basically one obtains their eq.(16) in the form:

$$T_{k'}(t) \sim \left(\exp(-i\mu v^2(t-t_o)/2\hbar)\right) \times$$

$$\int dk' A(k') \int dr \Phi(r,t) \exp(i k \bullet (r-vt)) \quad (14)$$

The introduction of intermediate I-frames conduces directly to semi-classic models. This form of the scattering cross section is well known.[20]

Without entering in details, the key point here corresponds to the mapping from an abstract space via partites' inertial frames towards real laboratory space. The passage from eq.(12) to eq.(14) is underlined in the present framework though not demonstrated.

**Probing: mapping a wave function?**
Going to a weak measurement scheme consider the partite wave function $|\Phi\rangle$ and the operator $T_R$ given as $|R\rangle\langle R| = \hat{\pi}_R$ followed by a strong momentum probing concerning the I-frame state. We are now referring to laboratory

space where I-frames origins are properly defined yet we formally follow Lundeen et al. approach [21]:

$$\langle k|\hat{\pi}_R|\Phi\rangle / \langle k|\Phi\rangle = \langle k|R\rangle\langle R|\Phi\rangle/\langle k|\Phi\rangle \quad (15)$$

This stands for the weak probing at R projected on k-space:

$$\langle \pi_R \rangle_{Weak} = \langle k|R\rangle\langle R|\Phi\rangle /\langle k|\Phi\rangle \rightarrow \exp(ipR/\hbar)\, \Phi(R) /\langle p|\Phi\rangle$$

Here, $k=p/\hbar$. Take Lundeen et al. case (page 188 ref[21]) to get

$$\langle \pi_R \rangle_{Weak} = \beta\, \Phi(R) \quad (16)$$

Thus, the weak probing extracts a wave function value in a neighborhood of real space point. While clearly, it is understood that $\langle p|R\rangle$ is modulated by a function peaked at a point (or region). Taking $\langle p|\Phi\rangle$ at p=0, then $\beta= 1/\Phi(R=0)$ is a constant absorbed in normalization.

The scheme is based on I-frames identification. The wave function $\Phi(R)$ hides all internal degrees of freedom that have to be explicitly analyzed.

The interesting point again is the role played by the quantum state in measurements processes. Because this is not an object, the role played by a quantum state is difficult to describe in common terms; there is no story on the materiality whereabouts no trajectories but probing. Equation (16) tells us that it will be the wave function projected in real space what is to be flagged out by the probing.[5] If the function value is zero there, it will be no response to the probe. And, again, a quantum state is not an object located in real space; yet *probing results* are. These latter are modulated only by wavefunctions via functional forms. Thus, *quantum states modulate probing responses*. The extrapolations leading to classical "representations" of particles is just a cultural avatar.

The amplitudes in the superposition, belongs to abstract space, modulate relative response and, when probed in intensity the square modulus as a whole is sensed. Remind a $T^\pm$ successful probe "destroys" the wave function of the one-system at a particular location: actually, it reshuffles the amplitude set.

Yet something is missing when laboratory level must be sutured to the abstract one. So far the wave function (e.g. $\Phi(R)$) covers the I-frame information as well as the "jumps" during excitation, yet they are not well characterized.[7] The energy field coupling these states providing energy information must be incorporated, otherwise the standard QM framework looks incomplete. This is the role of quantized electromagnetic fields.

Here, the photonic element must be properly handled; we move now to explicitly include this central element.

**Photon-matter framework**

The suture of abstract and laboratory spaces [1] requires completion of the standard QM scheme. The nature and origin of material supports are what labels distinguish in identifying base sets (kets and bras).

Photonic bases obtain that are useful to describe quantum states endurance under high intensity radiation and/or electrically charged surrounding elements. The nature of the quantum state stands out as key difference with previous approaches dominated by semi-classic models. At laboratory level, basic material elements (e.g.

fixed numbers of electrons and/or nuclei) *sustain* the quantum states (q-states); yet these q-states do not represent in a classical physics sense those objects. Q-states subsume information to be experimentally gleaned and/or modulated overcoming the idea of particles occupying (base) states;[5,6] actually, all possibilities accessible to a given system must be reckoned by the basis. Representation by particles is a characteristic of classical physics systems; such mode is banned from the present framework and replaced by the concept of quantum states *supported* (sustained) by a given materiality [1,5].

**Photonic basis**

Photon basis sets are identified by labels $n_\omega$, integers indicating number of EM energy quanta at frequency ($\omega$) that *can be exchanged* with quantized matter states; this statement defines a photon, certainly, not as a particle but as q-fields; Cf. Chpt. 5 in ref. [6], also ref. [22-24]

Suffices to note a base state symbol becomes: $|n_\omega\rangle$ or else $|n_\omega$; *plus other necessary labels*$\rangle$.

The (colored) vacuum signaled as $|n_\omega=0\rangle := |0_\omega\rangle$ indicates no energy quantum *available* for exchange at frequency $\omega$; yet, the corresponding lower *energy level* is: ½$\hbar\omega$! An energy level is just a label, not a quantity of energy at disposal; it is actually an element of a graded Hilbert space.

*Only differences of energy levels can be mapped to quantized energy.*

**Photonic Basis and Entanglement**

Photon base states and matter-sustained base states appear here combined in two guises: 1) direct products; 2) entangled bases sets elements.[1]

For a resonance case a two-state unit base derives from these elements. The set corresponds to what we call *photonic base state unit*.

Take a ground and excited states basis $|\varepsilon_k\rangle$ and $|\varepsilon_{k'}\rangle$: $\varepsilon_{k'}-\varepsilon_k = \hbar\omega$; actually one should use $\omega_{k'k}$ but it is too cumbersome.

The photonic base state unit reads:

$$(|\varepsilon_k\rangle \otimes |n_\omega=1\rangle \quad |\varepsilon_{k'}\rangle \otimes |n_\omega=0\rangle \ldots$$

$$|\varepsilon_k;1_\omega\rangle \quad |\varepsilon_{k'};0_\omega\rangle) \qquad (17)$$

The first two elements correspond to a bipartite case providing possibilities to describe photon absorption/emission processes that can eventually be mapped to real space laboratory level, e.g.: $|\varepsilon_k\rangle \otimes |1_\omega\rangle$ and $|\varepsilon_{k'}\rangle \otimes |0_\omega\rangle$.

The entangled base state elements $|\varepsilon_k;1_\omega\rangle$, $|\varepsilon_{k'};0_\omega\rangle$ contain information on energy gaps signaling entanglement of photon states and material field states: "photon dressed" base states. This notation emphasizes the relative character energy levels have. [1]

None of the basis elements entering the base can be handled as independent; a particle model introduces disjoint-ness thereby breaking quantum character.

If the photon energy does not match an excited energy level the entangled state correspond to an off-resonance case.

The subset under resonance condition presents the characteristic of sharing the same energy level value; i.e. at $\varepsilon_{k'=0}+\hbar\omega$ the generic four levels show thence energy degeneracy. This information is fully obliterated in the standard formalism.

Consider an energy band that can be accessed from diverse root states by choosing adequate external photon energies. Intra-band energy gaps being much smaller than average excitation energies. The characteristic frequencies within the selected band gaps may be found in the microwave and/or radio frequency (rf) radiation. In other words, after excitation putting amplitudes at the intra-band levels these states can be coupled via low frequency radiation to prompt time evolution within what may be called transition states bands.

**Electronuclear photonic basis**

Generalizing the standard molecular view to chemical systems as shown below one can use the preceding analyses and give a particularly interesting case by using Glauber coherent states.[22,23]

Quantum numbers for EN base states form non-separable products of an electronic j and subsidiary nuclear-related excitation quantum numbers k(j), this latter serves to identifying EN-excitation levels:

$$|j\ k(j)\rangle \Leftrightarrow |\varepsilon_{j\ k(j)}\rangle.$$

In [2-4,8,12-14] (and references therein) EN states are examined in detail.
According to eq.(17) photonic base states $|jk(j)\rangle \otimes |n_\omega\rangle$ and $|jk(j);n_\omega\rangle$ come in pairs: a direct product an a *photon entangled* basis elements; the present example incorporating the equivalent to $n_\omega$ photons; also $n_\omega = 1_\omega + \ldots + 1_\omega$ n-times stand for different possibilities. Off-resonant situations can also be included.

A resonant electronic excitation relates energy levels differing in electronic label: $|jk(j)=0\rangle \otimes |0_\omega\rangle$, $|j'k(j')=0\rangle \otimes |1_\omega\rangle$, the energy gap $E_j - E_{j'} = \hbar\omega$ signals a possible process in the j→j' direction, the system increases in one photon number; while in the j'→j direction the system decreases in one photon number; here the label j' signals a lower (ground) state with respect to j-label; 0-0 transition relates k(j)=0 to k(j')=0. "Jumps" appear only if one focus on either component separately: photon field or matter field. Otherwise, there are no jumps. The jumps are not physical phenomena they help describe particular cases.

The simplest case: photonic base states $|j'k(j')\rangle \otimes |1_\omega\rangle$ and $|jk(j)\rangle \otimes |0_\omega\rangle$, if j' is the ground level for j-level then at resonant frequency these photonic base states are energy degenerate and can be set to sustain a coherent state by employing another time dependent source manipulated by external agents (e.g. high intensity rf radiation); the direct product states become locked by this new low frequency EM radiation.

To indicate photon "absorbed", i.e. energy that is no longer accessible for exchange, the base states: $|j'k(j')=0; 1_\omega\rangle$, $|jk(j)=0;0_\omega\rangle$ help denote the situation; only base state $|j'k(j')=0\rangle \otimes |1_\omega\rangle$ permits handling incoming or outgoing photon states. More generally, for $|j'k(j')\neq 0\rangle \otimes |1_{\omega'}\rangle$ the transitions are reckoned from EN levels (Raman spectra for instance). The four elements in eq.(17) have then

the same energy value when referred to the same origin.

Off-resonance case: the energy gap $E_j - E_{j'} = \hbar\omega' > \hbar\omega$, the probe photon energy is smaller than the assigned gap $\hbar\omega'$. The base states $|j'k(j')\rangle \otimes |n_\omega\rangle$ and $|j'k(j');n_\omega\rangle$ single out a particular EN energy level with different energy label. The entangled base state shows the energy level $E_{j'}+\hbar\omega$ lying inside the gap $E_j-E_{j'}$ (by construction $\omega'>\omega$). The photonic state evolves in this subspace until getting reemission in an arbitrary direction that may differ from the incident direction of the incoming photon state that originated the situation: this would characterize an elastic scattering process.

**Quantum photonic states**

In what follows organize basis sets into "sectors" of un-entangled followed by the sector of entangled forms:
$$(\ldots |j'k(j')\rangle \otimes |1_\omega\rangle \ |jk(j)\rangle \otimes |0_\omega\rangle \ldots$$
$$\ldots |j'k(j');1_\omega\rangle \ |jk(j);0_\omega\rangle \ldots) \quad (18)$$

To check usefulness let introduce descriptions of several physical situations starting from the general state vector case:
$$(\ldots C_{j'k(j')\otimes 1\omega} \ldots C_{jk(j)\otimes 0\omega} \ldots$$
$$C_{j'k(j');1\omega} \ldots C_{jk(j);0\omega} \ldots)^T \quad (19)$$

The base state label is didactics.

Thus, the quantum state $(\ldots 1_{j'k(j')\otimes 1\omega} \ldots 0_{jk(j)\otimes 0\omega} \ldots 0_{j'k(j');1\omega} \ldots 0_{jk(j);0\omega} \ldots)^T$ seen from laboratory perspective might either correspond to ingoing or outgoing photon states label by the material I-frame.

Coherent states such as $(\ldots 0_{j'k(j')\otimes 1\omega} \ldots 0_{jk(j)\otimes 0\omega} \ldots C_{j'k(j');1\omega}(t) \ldots C_{jk(j);0\omega}(t) \ldots)^T$ do not elicit free photon available, it is an entangled photon-matter state.

When all base states of a given sector share the same I-frame, it is possible to describe them with Schrödinger-like time evolution involving full class photonic space. Again, the model system is not going to release a photon state.

Consider for instance a process given by the sequence below:

$$(\ldots 0_{j'k(j')\otimes 1\omega} \ldots 0_{jk(j)\otimes 0\omega} \ldots C_{j'k(j');1\omega}(t) \ldots$$
$$C_{jk(j);0\omega}(t) \ldots)^T \rightarrow (\ldots \exp(i\mathbf{k}\cdot\mathbf{x})1_{j'k(j')\otimes 1\omega}$$
$$\ldots 0_{jk(j)\otimes 0\omega} \ldots 0_{j'k(j');1\omega} \ldots 0_{jk(j);0\omega} \ldots)^T \quad (20)$$

This stands for a possible spontaneous photon emission from the I-frame system in direction $\mathbf{k}$. Thus, a detector located along that specific $\mathbf{k}$-direction at the tip of a well-defined $\mathbf{x}$ *may* interact with this quantum state in the guise of a correlated photon *state* (not a particle!).

The situation just described marks out a possibility only. For an abstract photon process case, $\mathbf{k}$-directions are infinite in number. Only one can show the photon energy (if the case is for one photon) via appropriate interaction operators.

Thus, if a "click" obtains it means presence at the detector of the state eq.(20). Though in this case, the event (click) reveals not only presence of a materiality in the form of EM-energy but also contains further information concerning the q-state; yet a "click" does not show up this explicitly. One cannot take a photography using one-photon energy only! Yet a full picture in terms of possibilities is there.

At the origin of vector $\mathbf{x}$ (I-frame) the material system can show response from the root state component $|j'k(j')\rangle$

as it displays a non-zero amplitude so it can be probed with another set of frequencies to finally identify it. A click is hence not a full characterization of the q-state.

In short: a click is not only a counted click. There is more to it. Thus, to reduce measurement to counting "dissipates" information as it were.

In this example all the remaining possibilities affecting the state vector have zero amplitude, the system state appears as a projection from the global one onto the non-entangled sector.

Incidentally, if a second photon state were to impinge over the system eq.(20), an induced photon state emission can occur. The intensity would increase with a plus signaling coherence.

In eq.(20) if amplitudes differing from zero simultaneously cover entangled and direct product base states (same I-frame), the quantum state is referred to as coherent q-state and as a function of time the amplitudes appear to be tied up. The coherent state responds as a whole; spontaneous emission events randomly happen but the states in average will show particular lifetimes.

This latter type of quantum process would characterize MW or rf radiation as catalyst to the extent the wave packet "transport" amplitudes from a reactant to a product channel state. The bandwidth may be in (quasi) resonance with a number of electronuclear states associated to different lower energy states. Time evolution within the band domain would put non-zero amplitudes at base states that are not accessible from the initial root state.[1]

This state of affair renders a system susceptible to external probing (modulators) at frequencies different from the one used in excitation from ground state.[1]

In summary, the photonic base set incorporates all basis states that might be activated from external sources (amplitude modulations) and consequently be able to sustain interactions with quantum physical probes; eventually those showing non-zero values can be directed as roots for measuring or inducing signals. The quantum states sitting at the detector device must be able to sense most of the photon states involved by the system under measurement and translate them to signals interpretable by an experimenter (eventually an observer though this latter is a literary figure of the past) or communicate to other quantum system in a network. Probing of q-state left behind by the emitted photons is another possibility; a sort of pump-probe situation.

**Coherent/Decoherent States**

The problem of interactions between a coherent quantum state with macroscopically distinguishable state systems is a first step to map out micro-to-macroscopic levels. A simple model is used here to consider some qualitative aspects.

Take the equivalent to eq.(20), but now let us use the coherent state known as "Schrödinger-cat" states;[23] the basis interaction unit reads:

$$(\ldots|\varepsilon_{k'=0}\rangle \otimes |A_\omega\rangle \ldots |\varepsilon_{k'}\rangle \otimes |0_{A\omega}\rangle \ldots$$
$$|\varepsilon_{k'=0}; A_\omega\rangle \ldots |\varepsilon_{k'}; 0_{A\omega}\rangle \ldots)$$
(21)

The first term would appear either as incoming states onto a material system sustaining an EN base states or as outgoing quantized EM field. The second term, the photon base state stands for complete depletion so that implicitly that energy would be coherently stocked by materiality. Re-emission q-state the amplitude would be multiplied by a k-factor.

Entanglement possibility concerns q-states of the form:

$(\ldots 0_{k'=0\otimes A\omega}\ldots 0_{k'\otimes A\omega}\ldots C_{k'=0;A\omega}\ldots C_{k';0\omega}\ldots)$. (22)

That is a linear superposition sustained by entangled base states. While a state given as:

$(\ldots C_{k'=0\otimes A\omega}\ldots C_{k'\otimes A\omega}\ldots C_{k'=0;A\omega}\ldots C_{k';0\omega}\ldots)$,

where all amplitudes show non-zero values would correspond to a non-separable photon-matter q-states. In both cases there will be no accessible free photon states.

Interestingly, to activate a quantum process involving new possibilities would require external low frequency radiation whenever other j-states were accessible from the coherent state via intermediate decoherence states. Different j-label relate to possible lower energy temporal stable states (dark states for example).

This type of processes correspond to chemistry from "above" that femto- and atto-second pulses might open.

*Decoherence* would show up if the q-state below is seen as source of photon "flash":

$(\ldots 1_{k'=0\otimes A\omega}\ldots 0_{k'\otimes A(n-1)\omega}\ldots 0_{k'=0;A\omega}\ldots 0_{k';0\omega}\ldots)$.

In other words, it would stand as emission of a coherent photon bunch originated at the corresponding I-frame or alternatively a colored pulse.

The vector state:

$(C_{k'=0\otimes A(n-1)\omega}\ldots C_{k'\otimes A(1)\omega}\ldots C_{k'=0;A(n1)\omega}\ldots C_{k';1\omega}\ldots)$

known as Schrödinger-cat state holds coherence since the state with one photon less is not distinguishable for large n;[6] for situations where high power lasers dominate, such would be the case.

If normalized vector takes on the form $C_{k'=0;A\omega}$ = sin ν and $C_{k';0\omega}$ = cos ν one obtains the linear superposition state covering the entangled sector only:

|φ$_A$> = cos ν |j'k(j');A$_\omega$> + sin ν |jk(j);0$_{A\omega}$>    (23)

This state can be submitted to photo-detection and homodyning. As shown by Teo et al[25] such q-state is useful to exploring deep questions at the Fence [5,6] concerning realistic loophole-free Bell test states with atom-photon entanglement.[22-25] This theme is of great import though we leave it now; the key issue is that macroscopic photon states can be used within the present framework constituting a resource to exploit in future studies.

Important analyses of EN-EM field interactions can be found in refs.[20,25] Decoherence can be located here as a process where amplitudes and I-frames independence involve the direct product base states. Detection of a fragment that does not recall information of the remaining ones can be seen as de-coherent response.

Before moving on, notice two things: 1) time evolution can take place with photon-dressed states, namely photon-matter entangled; 2) Any event incorporating photon emission requires a minimal subspace where all generic possibilities are considered (entangled and non-entangled).

**Apparent Photonic "2-states" model**

The two sectors of photonic base (simplified) in a one photon case looks like:

(…|j'k(j')>⊗|1$_\omega$>…|jk(j)>⊗|0$_\omega$>…
|j'k(j');1$_\omega$>…|jk(j);0$_\omega$>…)    (24)

Examine first a drastic reduction of this base vector by deleting direct product components. The model simplify the writing down to:
→ (…|j'k(j')>…|jk(j)>…)      (24')
Thereafter, a second model type obtains: retention of non-entangled photon-matter base states sector, thus eliminating entangled base states. In this case one is back to quantum jumps picture. Lost is the possibility to see the effect of microwave radiation for instance in chemical change. But it is useful to gain some experience with this type of approach first.

Suppose before level crossing the external label reads $\mathcal{F}_>$ and after this $\mathcal{F}_<$, just conventionally. At label $\mathcal{F}_>$-region energy $E_{j'k(j')=0}(\mathcal{F}_>) < E_{jk(j)=0}(\mathcal{F}_>)$ while at $\mathcal{F}_<$ the energy $E_{j'k(j')=0}(\mathcal{F}_<) > E_{jk(j)=0}(\mathcal{F}_<)$; a sort of X-form; energy levels orders are inverted by the field $\mathcal{F}$ external to the model system.

Suppose one can design a source that $E_{j'k(j')=0}(\mathcal{F}_>)$ stepwise increases so that one gets a sequence:
$E_{j'k(j')=0}(\mathcal{F}_>)$, $E_{j'k(j')=1}(\mathcal{F}_>)$, $E_{j'k(j')=2}(\mathcal{F}_>)$,…, $E_{j'k(j')=n^*}(\mathcal{F}_>)$,…      (25)
This models an excitation path without change of electronic label type; a model vertical excitation path; a second excitation operator is required with energies that may range in the microwave spectrum to move label k(j'). The value of n* is chosen so that the energy level comes to a neighborhood of the value $E_{jk(j)=0}(\mathcal{F}_>)$.

Once the energy gap is filled with the appropriate number of EN-excitations via external low frequency photons, the two energy levels $E_{j'k(j')=n^*}(\mathcal{F}_>)$ and $E_{jk(j)=0}(\mathcal{F}_>)$ approach each other so they might be almost degenerate. At least the following bracketing of the j-level holds:
$E_{j'k(j')=n^*-1}(\mathcal{F}_>) < E_{jk(j)=0}(\mathcal{F}_>) < E_{j'k(j')=n^*+1}(\mathcal{F}_>)$
(26a)
At ($\mathcal{F}_<$), mutatis mutandis, one can get:
$E_{jk(j)=n^*-1}(\mathcal{F}_<) < E_{j'k(j')=0}(\mathcal{F}_<) < E_{jk(j)=n^*+1}(\mathcal{F}_<)$
(26b)

So far the model has no mechanism to mix j and j' labeled states. To mix j and j' levels let introduce external couplings λV(t) effecting electronic transition amplitudes:
<jk(j)| λV(t)|j'k(j')> ≠ 0.

The resulting coherent electronic states |+>=|E$_+$> and |->=|E$_-$> change with the external field label as calculated above. But now, physical mechanisms are emphasized to describe transitions among them.
$E_+(\mathcal{F})$-$E_-(\mathcal{F})$ is minimal at crossing of bare energy levels determining an anti crossing zone. And, while the gap is always positive at finite field, the energy gaps for λV(t)≠0 would model change.

Taking advantage of photonic base sets let us examine this issue in more detail. Suppose the system has been prepared by adding a certain number of excitations n*. Consider excitation at ($\mathcal{F}_<$) engaging levels j'→j
(…$C_{j'k(j')=n^*;0\omega'}(t)$…$C_{jk(j)=0;1\omega'}(t)$…)$^T$

Prepare the system in the state:

$$(\ldots C_{j'k(j')=n^*;0\omega'}(t_o)=1\ldots C_{j\,k(j)=0;1\omega'}(t_o)=0\ldots)^T$$

(27)

If no coupling is effective this initial state stays put.

Thus: How these amplitudes can be modulated?

For the time being examine a change the amplitudes prompted by an external effector:

$$(\ldots C_{j'k(j')=n^*-1;0\omega'}(t_1)\approx 1\ldots C_{j\,k(j)=1;1\omega'}(t_1)\approx 0\ldots)^T$$

(27a)

We describe this as an apparent excitation transfer of one low frequency photon from the j'-ladder to the j-ladder; continuing this process until getting e.g.:

$$(\ldots C_{j'k(j')=n^*-m;0\omega'}(t_m)\ldots C_{j\,k(j)=m;1\omega'}(t_m)\ldots)^T$$

(27b)

One ends at state looking like:
$$(\ldots C_{j'k(j')=0;0\omega'}(t_m)\ldots C_{jk(j)=m;1\omega'}(t_m)\ldots)^T .$$

(27c)

We choose it to have the same energy as
$$(\ldots C_{j'k(j')=0;0\omega}(t_m)\ldots C_{jk(j)=m;1\omega'}(t_m)\ldots)^T.$$

(27d)

Relaxation at j-channel would produce a response from ground state related to j-level.

Note that the "transferred" photon frequency differs markedly from ω' that would correspond to an electronic 0-0 transition. [1]

If this transfer process can proceed, the q-state would look like:
$$(\ldots C_{j'k(j')=1;0\omega'}(t_{n^*-1})\sim 0\ldots C_{jk(j)=n^*-1;1\omega'}(t_{n^*-1})\sim 1\ldots)^T$$

(27e)

This is a state in a neighborhood of total low frequency energy transfer. The amplitude "localizes" at base state |jk(j)=n*-1; ω'>; observe again that ω' label the state giving us a particular information.

For a symmetric model the excitation frequency ω' is the same at both extremes separated by the crossing.

The conversion of j'-colored low frequency excitation (at fixed ω') into j-colored one puts in evidence non-separability of electronic and nuclear excitations.

Let now write down the "product" state in chemical sense given here as a change of principal (electronic) quantum number:

$$(\ldots C_{j'k(j')=0;0\omega'}(t_{n^*})=0\ldots C_{j\,k(j)=n^*;1\omega'}(t_{n^*})=1\ldots)^T$$

(27f)

The label $t_{n^*}$ identifies full transfer of colored photons for this model. In the amplitude $C_{jk(j)=n^*;1\omega'}(t_{n^*})=1$ the label points either to a parentage with j' label (1ω') while $E_{jk(j)=n^*}$ points to EN level of the electronic state j. A chemical change (change j' to j) is mediated by EN excitations. The "passage" is made above the crossing without geometry change controlled by a mechanical "reaction coordinate". Interestingly, there is no "conical intersection" involved; only quantum physical processes [1].

The transfer can be seen as a sequential inversion of the quantum state at constant energy-label $E_+$.

Transport from one "side" to the other (abstract X-form) corresponds to an EN wave packet propagation where low frequency excitations drive the passage process that cannot be unless labels j and j' are coupled. State (27e) is equivalent to the one that is to be found at the extreme of the X-arrangement above. The coupling of low frequency transitions leading to energy transfer between the j-and j'-ladders would exemplify a chemical transformation as well as a conformational change; see ref.[1]

Pictorially speaking, transfer of "colored" excitations screen the "external field" $\mathcal{F}_>$ until getting the effect that would correspond to reversing it into $\mathcal{F}_<$; hidden there, is a polarity reversal sustained by the materiality. This is a way of speaking that may help describe the simulation.

A transition state zone suggests itself, as watermark, yet this time with no potential energy functions; photonic-states spectral systems are there to sustain such process. And the same (*invariant*) *elementary materiality sustains* the chemical change that is expressed by the electronuclear quantum numbers appearing as labels to relevant amplitudes.

**Feshbach resonances**

These states characterize interaction of bound with states in the continuum.[4] In terms of multipartite systems these interactions relate systems differing at least in one direct product forms. Basically, in terms of I-frames both states present different numbers of nodal planes.

The mapping relating q-states eq.(9) and (10) would include, as a particular case Feshbach resonant states.[4]
States belonging to the n+1-partite set may belong to two classes: 1) mix of uncorrelated fragments' states; 2) Entangled states coming from n-partite sector.

Thus, knitting together states from two separated I-frames (n+1-partite) bringing them to an I-frame (n-partite) requires that these states must be orthogonal to the n-partite ones.

In this respect, a couple of points are worth pointing out:

1) Energy scales origins differ between n- and n+1 partite base states; for the latter the lowest "dissociation" energy is taken as the energy sum of two ground asymptotic states;
2) The n-P system would show energy levels positive and negative with respect to the asymptotic value;
3) Relative kinetic energy (RKE) may displace the location of (n+1)-P levels compared to n-P one;
4) Energy gaps labels must consider this situation;[1]
5) For given RKE the level sum in (n+1)-P can be put in near degeneracy with some levels belonging to the n-P;
6) Uncorrelated fragments lack definite global $S^2$;
7) All q-states of n-P asymptotically correlated to n+1-P can be proper states of a given $S^2$.
8) For multi-partite elements, total angular momentum would depend on the number of nodal planes. For fragments independently put together both $S^2$ and $L^2$ might not provide good quantum numbers, L stands for space angular momentum. Thus, reactivity of "free radicals" is a concept to be examined carefully on a case-by-case ground. For instance, two ground state hydrogen atoms do not react unless a third body acts as catalyst.[2]

The photonic approach permits natural activation mechanisms. Yet, knitting non-commensurable systems require further elaboration; other aspects are examined below.

## Attosecond pulses

Attosecond duration pulse is a relatively recent tool available to study EN systems where the electronic quantum number plays central roles. These pulses must have high time resolution to actually explore the photon assisted quantum states. High-order harmonic-base sources delivering pulses of extreme ultraviolet (XUV) radiation with duration in attosecond range [27,28] generate inside the pulse low frequency photonic states (overtones).

An XUV pulse centered at a frequency $\omega_o$ targeting a region above ionization threshold of a given EN system covers a relatively large range of low frequencies above and below $\omega_o$. The pulse would overlap the continuum as well the highest excited levels sustained by a given materiality with radiation-quantized gaps in the low frequency range. The root state being among the lowest energy levels of the system under study the energy gap "activates" base states above a limit either of dissociation and/or ionization.

## Attosecond time scale processes

Control of a re-collision electron in a strong laser field does not provide an appropriate quantum physical picture.

At the energy range where ionization and dissociation processes dominate the chemical concepts of atom and molecules are not much relevant. It is not the mechanical electron-ion photo recombination that generates attosecond optical pulses; it is the coherent quantum state with finite lifetime that might be origin. Thus, emission of (free-) electron states following interaction with an attosecond pulse cannot be an instantaneous process. One has to avoid the trap opened by a semi-classic calculation; time involved is not commensurate to lifetimes.

## $H_2$ & $H_2^+$ & $e^-$ Partite base set

There are four generic sets: 1-partite $|H_2>$, bi-partite $|H_2^+> \otimes |e^->$, tri-partite bases $|H> \otimes | H^+> \otimes |e^->$ and quadric-partite $|H^+> \otimes | H^+> \otimes |e^-> \otimes |e^->$. There is a subset for ion molecular hydrogen containing bases for $|H_2^+>$, $|H> \otimes |H^+>$ and $|H^+> \otimes | H^+> \otimes |e^->$. Disregarding the connection of $H_2^+$ with $|H_2>$ one may discuss only the latter subset.

Inclusion of configuration space and more detailed labeling with quantum numbers like those for eqs.(1-1)-(1-m) allow detailed analyses:

$|H_2^+>$: $<X_1,X_2,x_3|\phi_{k'1}\phi_{k'2}\phi_{k'3}>$; 1-P    (28-1)

$|H^+> \otimes |H>$: $<X_1|\phi_{k'1}> \otimes <X_2,x_3|\phi_{k'2}\phi_{k'3}>$: 2-P
    (28-2-1)

$|H> \otimes |H^+>$: $<X_1,x_3|\phi_{k'2}\phi_{k'3}>) \otimes <X_2|\phi_{k'1}>$; 2-P
    (28-2-2)

$|H^+ H^+> \otimes | e^->$: $<X_1,X_2|\phi_{k'1}\phi_{k'2}> \otimes <x_3|\phi_{k'3}>$; 2-P
    (28-2-3)

$|H^+> \otimes |H^+> \otimes |e^->$: $<X_1|\phi_{k'1}> \otimes <X_2|\phi_{k'2}> \otimes <x_3|\phi_{k'3}>$; 3-P    (28-3)

Generic quantum states covering the complete base set read as:

$|\Omega,t> \Rightarrow$
$(|28-1> ... |28-2-1> |28-2-2> |28-2-3> ... |28-3>) \bullet$
$(C_{28-1}(\Omega,t) ... C_{28-2-1}(\Omega,t) \; C_{28-2-2}(\Omega,t) ...$
$\quad C_{28-2-3}(\Omega,t) ... C_{28-3}(\Omega,t))^T$    (29)

A broadband attosecond pulse width may activate matter-sustained excited states around an energy range above a photoemission threshold. The resulting coherent quantum state would necessarily present a lifetime; hence there will be a time delay before some decoherence channels effectively show up. The phenomenon of delay in photoemission can originate from activation of such coherent quantum state.

Technically, an XUV pulse would activate a photonic-unit-base-state that reads now like eq.(21); this time one gets a bunch of different low frequencies.

As discussed above for EN base states, a manifold opens up to the system above and below the central frequency carried by a pulse; these are higher harmonic excitations rooted at the EN excitation spectra. Thus, what could evolve after the attosecond time scale excitation is a q-state packet driven by the high harmonics generated with high intensity and low frequencies EM radiation. An alternative is to shine low frequency radiation: e.g. rf radiation and even lower to modulate time evolution.

One clear difference with a semi-classic picture (free electron vibrating with the EM electric field) is the emergence of entangled EM/EN quantum states (coherent linear superpositions). As the amplitude over the corresponding base state takes on non-zero values, no free electrons are accessible. The coherent state may show characteristic lifetimes and free-electron states traceable to decoherence.

In other words, if an electron ejection would happen it will be delayed by the entangled contribution to the quantum state.

**Attosecond pulse probing**

Experiments related to $H_2^+$ actually start up with di-hydrogen system ($2e^-$ and $2H^+$). 1-partite base states $<x_1,x_2,x_3,x_4|\phi_{k'1k'2k'3k'4k'}>$ include ground and all excited states of the hydrogen molecule. Electronic excited above $H_2$ ionization base states interact with EN states of $H_2^+$ and to get a first ionization particular attosecond pulses can achieve preparation.[26]

A key method of synthesizing atto-second pulses was the generation, in rare gases, of high harmonics (HH) of an infrared femtosecond laser pulse. For instance, high orders (300 or more) coherent harmonics may cover spectral bandwidths of hundreds of electron volts. Technology has advanced to produce relatively narrow wave packet that include few HH states around maximum peak [25]. See Dahlström et al.[28] and references therein for technical descriptions.

Thus a coherent attosecond pulse would set up a coherent photon assisted state where the amplitudes would cover a broad range; such non-zero amplitudes may extend over ionization/dissociation bases. To the extent that excitation is coherent the resultant quantum state would share coherency and, as a result, there will be no free-particle states accessible. This includes non-free photon states. There will necessarily be a delay in breaking coherence thereby leading to *delayed photoionization* products.

The photonic approach emphasizes the role of coherent states resulting from laser pulses probing. A new kind of chemical process: a photonic one displays in front of us implying that emission must show finite lifetime. There is no "free" electron dance in the laser field. Mixing of quantum to classic descriptions does not help much in this case.

In particular, the photonic approach indicates that no semi-classic EM field is left active in the resonance framework. This is the price taken by non-separability. Thus to proceed in laboratory space there is need for external EM fields such as free microwave and radio frequency radiation to modulate time evolution.

This field is at an earlier development phase, yet laser assisted chemistry is in a vigorous development. We hope this paper may be useful to chemists, biologists and quantum technologists in this respect.

**Discussion**

A fully quantum physical photonic framework presented here combines matter and photon fields excitations, yet leaving room to incorporate results and ideas rooted at advanced quantum chemical schemes. [1,3,4]

The paper addresses some basic problems in quantum physics and quantum mechanics in particular. In opinion of Steven Weinberg there is no entirely satisfactory interpretation of quantum mechanics.[29] That, in a way Hobson's [30] work corroborates with a very interesting analysis of two-photon interferometry thereby discussing quantum state collapse in a manner differing from the standard view; yet grounding elements such as particles and representational mode remain in widespread use. This is surprising since the question rose by Laloë [31] asking whether one really understands quantum mechanics still remains unanswered.

The present work and those reported in[1,5,6] credit the idea that we are rather far from understanding quantum mechanics if we insist using classical physics tenets; the oxymoronic character of wave-particle duality is difficult to hide.

The ideas presented here radically differ from the dominating archetype. The discussion of entanglement presented above illustrates this point. Entanglement concerns q-states not the elementary material system sustaining it. Consequently an interpretation of quantum mechanics, if that one is to refer to the materiality sustaining quantum states, becomes dispensable.

Yet, *event* sequences as such, e.g., those recorded under laboratory premises, can be submitted to probabilistic analyses; they share an "objectiveness" character (I-frame sustained). In this case the weight is put on recorded material: the spot (click). In this context it is reasonable that Bayesian analyses [32] turn out to be useful since the enquiring direction is from outside (laboratory) to the inside (quantum). However, the supporting role of materiality with respect to quantum states, that is fundamental for the present view, fades away in a probabilistic (counting) context. And, consequently quantum mechanics' weirdness's would show up again.

To bypass such contradictory way of speaking we ought to accept that Classical physics and Quantum physics languages are, strictly speaking, irreconcilable. Yet they can still be used in an harmonizing manner with the classical mode playing a subsidiary and pedagogic role; the introduction of inertial frames with partite base sets is one example.[1,3,4]

The multipartite and multi-I-Frames scheme widen applicability of quantum approaches in a variety of directions:

1: It might be adapted to produce a QM without wavefunctions following work by Schiff and Poirier.[33] Thus, simulation procedures balancing abstract with semi-classic quantum schemes[3,4] may shed new light on a number of physical-chemical-biological processes.

2: 3D imaging by Mass spectrometry to show partite distributions can be a powerful analytical tool.[34]

3: For polar partite, cooling down constituents with continuous centrifuge decelerators offer another possibility [35] to study partite elements in detail.[36]

4: Quantum-confinement using Stark effect e.g. semiconductor nano-elements developed by Weiss and coworkers,[37] provide further openings.

5: The concept of chemical bond translates into entanglement of partite q-states. Quantum correlations of partite states elicit in turn presence of chemical bond; no pictorial views.

These examples illustrate possibilities where the present quantum approach may found useful applications when photon fields are in action.[38-40]

Coherence is a key concept that becomes simply formulated in a multipartite photonic framework. The size of the materiality sustaining quantum states is not a restrictive element. The classic view is not apodictic though it might help us on the way to apprehending quantum phenomena. Clearly, it cannot help understand the quantum phenomenon itself; we believe the wave-particle duality has blurred key issues only.[5]

Resonance, on the other hand, between a discrete energy level with a continuum is just Feshbach sort of state. Dissociation towards *free partite-q-states* would take this path; although if entanglement, as discussed here, shows up then this would not be erased unless a demolition probing applies.

An irreversible laboratory process recorded by a measuring apparatus stands as one possible outcome among all possible ones; these latter are somehow "baked" into the physical quantum state; and collecting the events, according to present view, an "image" of the quantum state would emerge.[5,6]

Events belong to laboratory world; they elicit materiality in the first place. Yet they also carry q-state information that is only apparently "dissipated". It suffices collecting a sufficient number of these events to initiate portraying the q-state. This basically is the ground for the weak measurement scheme.

Also, the events can be used to set up clocking so that lifetimes to the system sustaining the wavefunction can be assigned. The photonic units bases provide the information while the amplitudes modulate the interactions.

The concept of photon-matter base state unit implements the basic hypothesis advanced at the beginning of this paper. It includes the photon state emission possibility as well as the opposite; amplitudes modulate the response. An external interaction yielding a photon emission view from an external I-frame system leads to entanglement of source state and a signal that can be detected and clocked by another system at laboratory premises. The laboratory time *is not* correlated to the time in internal space. This is a conundrum the solution of which is well beyond present author's capabilities to solve.

Furthermore, with parametric time-reverted q-states, absorption-emission phenomena can be simply integrated. Such inversion makes sense in abstract space (parametric time axis) only. In laboratory space we can wait for an echo (resurrection...).

The photonic quantum state plays the function of bridging abstract Hilbert space elements to the laboratory ones. *It has no representational character*.

A functional role for wave functions is hence apparent coming from the presentation given in the scattering framework and probing section.

Yet, ignoring the photonic side amputates such functional character to wave functions.

Last but not least, our paper touches issues raised by Anton Zeilinger in his beautiful essay: "The message of the quantum".[45] The reader can now contrast views and extract his/her own conclusions. A cross-analysis will be presented elsewhere.

**Acknowledgements**

The author thanks all coworkers that have helped develop these ideas. In particular Prof. R. Contreras (Chile University), G. Arteca (Laurentian, Sudbury, Canada), JM. Aulló (Valencia University) and E. Brändas (Uppsala University).